\documentclass[11pt]{article}
\usepackage{amssymb,fullpage}
\usepackage{times}
\usepackage{epsfig,theorem,fullpage,amssymb}
\newtheorem{theorem}{Theorem}
\newtheorem{lemma}{Lemma}[section]

\newtheorem{claim}{Claim}

\newcommand{\Xomit}[1]{}

\newcommand{\OPT}{\mbox{\textsc{opt}}}

\newcommand{\A}{{\cal A}}
\newcommand{\R}{{\mathcal R}}

\newcommand*{\be}{\begin{equation}}
\newcommand*{\ee}{\end{equation}}
\newcommand{\ba}{\begin{eqnarray*}}
\newcommand{\ea}{\end{eqnarray*}}

\newcommand{\eps}{\varepsilon}

\newenvironment{proof}{\noindent {\bf Proof $\;$}}
                    {\hfill$\square$}

\begin{document}

\title{Covering selfish machines}
\author{Leah Epstein\thanks{
Department of Mathematics, University of Haifa, 31905 Haifa,
Israel. {\tt lea@math.haifa.ac.il}. }
\and
Rob van Stee\thanks{Department of Computer Science, University of Karlsruhe,
D-76128 Karlsruhe, Germany. \texttt{vanstee@ira.uka.de}.
Research supported by the Alexander von Humboldt Foundation. } }
\maketitle

\begin{abstract}
We consider the machine covering problem for selfish related
machines. For a constant number of machines, $m$, we show a
monotone polynomial time approximation scheme (PTAS) with running
time that is linear in the number of jobs. It uses a new
technique for reducing the number of jobs while remaining close
to the optimal solution. We also present an FPTAS for the classical
machine covering problem (the previous best result was a PTAS)
and use this to give a monotone FPTAS.

Additionally, we give a monotone approximation algorithm with
approximation ratio $\min(m,(2+\eps)s_1/s_m)$ where $\eps>0$ can
be chosen arbitrarily small and $s_i$ is the (real) speed of
machine $i$. Finally we give improved results for two machines.

Our paper presents the first
results for this problem in the context of selfish machines.
\end{abstract}

\section{Introduction}
Internet users and service providers act selfishly and
spontaneously, without an authority that monitors and regulates
network operation in order to achieve some social optimum such as
minimum total delay. An interesting and topical question is how
much performance is lost because of this. This generates new
algorithmic problems, in which we investigate the cost of the
lack of coordination, as opposed to the lack of information
(online algorithms) or the lack of unbounded computational
resources (approximation algorithms).

There has been a large amount of previous research into
approximation and online algorithms for a wide variety of
computational problems, but most of this research has focused on
developing good algorithms for problems under the implicit
assumption that the algorithm can make definitive decisions which
are always carried out. On the internet, this assumption is no
longer valid, since there is no central controlling agency. To
solve problems which occur, e.g., to utilize bandwidth
efficiently (according to some measure), we now not only need to
deal with an allocation problem which might be hard enough to
solve in itself, but also with the fact that the entities that we
are dealing with (e.g. agents that wish to move traffic from one
point to the other) do not necessarily follow our orders but
instead are much more likely to act selfishly in an attempt to
optimize their private return (e.g. minimize their latency).

\vspace{10pt}
{\bf Mechanism design} is a classical area of research with many
results. Typically, the fundamental idea of mechanism design is
to design a game in such a way that truth telling is a dominant
strategy for the agents: it maximizes the profit for each agent
individually. That is, each agent has some private data that we
have no way of finding out, but by designing our game properly we
can induce them to tell us what that is (out of well-understood
self-interest), thus allowing us to optimize some objective while
relying on the truthfulness of the data that we have. This is
done by introducing \emph{side payments} for the agents. In a
way, we reward them (at some cost to us) for telling us the truth.
The role of the mechanism is to collect the claimed private data (bids),
and based on these bids to 
provide a solution that optimizes the desired objective, and hand out
payments to the agents. The agents know the mechanism and are
computationally unbounded in maximizing their utility.

The seminal paper of Archer and Tardos~\cite{ArcTar01} considered
the general problem of one-parameter agents.
The class of one-parameter agents contain problems where any agent
$i$ has a private value $t_i$ and his valuation function has the
form $w_i\cdot t_i$, where $w_i$ is the work assigned to agent $i$.
Each agent makes a bid depending on its private value and the
mechanism, and each agent wants to maximize its own profit.
The paper~\cite{ArcTar01} shows
that in order to achieve a truthful mechanism for such problems,
it is necessary and sufficient to design a \emph{monotone}
approximation algorithm. An algorithm is monotone if for every
agent, the amount of work assigned to it does not increase if its
bid increases. More formally,
an algorithm is monotone if given two vectors of length $m$,
$b,b'$ which represent a set of $m$ bids, which differ only in
one component $i$, i.e., $b_i>b_i'$, 
and for $j\neq i$, $b_j=b_j'$,
then the total size of the jobs (the work) that machine $i$ gets from
the algorithm if the bid vector is $b$ is never higher than if
the bid vector is $b'$.

Using this result, monotone (and therefore
truthful) approximation algorithms were designed for several
classical problems, like scheduling on related machines to
minimize the makespan, where the bid of a machine is the inverse
of its speed~\cite{ArcTar01,Archer04,AuDePP04,AnAzSo05,Kovacs05},
shortest path~\cite{ArcTar02,ElSaSt04}, set cover and facility
location games~\cite{DeMiVa03}, and combinatorial
auctions~\cite{LeOCSh99,MuaNis02,ArPaTT03}.

\paragraph{Problem definition}
In the current paper, we consider the problem of maximizing the
minimum load (cover) on related machines.
This is motivated by situations where a system is alive
(i.e.~productive) only when all the machines are alive.
Another example is a system that needs to spread the risks. Say the
machines are sub-contractors, and the auctioneer want to spread the projects
(jobs) between them evenly, to spread the overall risk. Such a system is
interested in using all the machines, while maximizing the least loaded
machine.

Denote the number of jobs by $n$, 
and the size of job $j$ by $p_j$ ($j=1,\dots,n$).
Denote the number of machines by $m$, 
and the speed of machine $i$ by $s_i$ ($i=1,\dots,m$). %
In our model, each machine belongs to a selfish user.
The private value ($t_i$) of user $i$
is equal to $1/s_i$, that is, the cost of doing one unit of work.
The load on machine $i$, $L_i$, is the total size of the jobs assigned to
machine $i$ divided by $s_i$. 
The profit of user $i$ is $P_i-L_i$, where $P_i$ is the payment to user $i$
by the payment scheme defined by Archer and Tardos~\cite{ArcTar01}.

Our goal is to maximize $\min_i L_i$.
This problem is NP-complete in the strong sense~\cite{GarJoh79} even
on identical machines.
In order to analyze our approximation algorithms we use the
approximation ratio. For an algorithm $\A$, we denote its cost by
$\A$ as well. An optimal algorithm is denoted by \OPT.  The
approximation ratio of $\A$ is the infimum $\R$ such that for any
input, $\A \leq \R \cdot \OPT$. If the approximation ratio of an
offline algorithm is at most $\rho$ we say that it is a
$\rho$-approximation.

\paragraph{Previous results (non-selfish machines)}
For identical machines, Woeginger~\cite{Woegin97} designed a
polynomial time approximation scheme (PTAS). He also showed that
the greedy algorithm is $m$-competitive. No deterministic online
algorithm can do better. Other offline approximation results are
given in \cite{DFL82,CKW92}. Azar and Epstein~\cite{AzEp97}
presented a randomized $O(\sqrt{m}\log m)$-competitive online
algorithm and gave an almost matching lower bound of
$O(\sqrt{m})$.

In~\cite{AzaEps98}, a PTAS was designed for related machines. For
the semi-online case in which jobs arrive in non-increasing order,
~\cite{AzEp97} gave an $m$-competitive algorithm called
\textsc{Biased-Greedy} and showed that no algorithm could do
better. The well known Least Processing Time (LPT) algorithm does
not provide finite approximation ratio; given
two machines of speeds 1 and 4, and two jobs of size 1, it will
assign both jobs to the machine of speed 4.
\textsc{Biased-Greedy} is
a special case of LPT which prefers faster machines in case of
ties. We can see that even this variant gives a relatively high
approximation ratio.  It is known that
LPT is
not monotone but an adaptation called LPT* is
monotone~\cite{Kovacs05}. However, the adaptation acts the
same on the above input and thus it cannot be used for the
monotone covering problem. Moreover, since \textsc{Biased-Greedy}
acts as LPT on some inputs, it cannot be expected to be monotone
either.

For the case where jobs arrive in non-increasing order and
also the optimal value is known in
advance,~\cite{AzEp97} gave a 2-competitive algorithm
\textsc{Next Cover}.

\paragraph{Our results}
%
We present a \emph{monotone} strongly polynomial time
approximation scheme (PTAS) for a constant number of related
machines. Its running time is linear in the number of jobs, $n$.

We also present an FPTAS for non-selfish related machines 
(the classical problem). We use this to give a monotone FPTAS with
running time polynomial in $n$ and $\eps$ and the logarithm of
sum of job sizes.

Additionally, we present a monotone approximation algorithm based on
\textsc{Next Cover} which achieves an approximation ratio of
$\min(m,(2+\eps)s_1/s_m)$. This algorithm is strongly
polynomial-time for an arbitrary number of machines. Given the negative
results mentioned above, it seems difficult to design a monotone
approximation algorithm with a constant approximation ratio for
an arbitrary number of machines.
Finally, we study two monotone algorithms for two machines, and
analyze their approximation ratios as a function of the speed
ratio between the two machines. These algorithms are very simple
and in many cases faster than applying the PTAS or FPTAS on two
machines.

\paragraph{Sorting}
Throughout the paper, we assume
that the jobs are sorted in order of non-increasing size
($p_1\geq p_2\geq \dots \geq p_n$), except in Section
\ref{sec:ptas}, and the machines are sorted in a fixed order of
non-decreasing bids (i.e.~non-increasing speeds, assuming the
machine agents are truthful, $s_1\geq s_2\geq\dots\geq s_m$).

\section{PTAS for constant $m$}
\label{sec:ptas}

This section is set up as follows. In Section
\ref{sec:scale}, we prove some
lemmas about the amount of different sizes of jobs. 
In Section \ref{sec:monotone},
we show how to design a constant time simple optimal monotone
algorithm for an input where the number of jobs is constant
(dependent on $m$ and $\eps$). In Section \ref{sec:reduce}, we
show how to reduce the number of jobs to a constant, allowing us
to find the optimal value for this changed instance in constant
time. In Section \ref{sec:optimal}, we show that due to this
reduction, the optimal value is reduced by at most
$\eps\cdot\OPT$. Finally in Section \ref{sec:time}, we show that
our algorithm has linear running time in the number of jobs.
Altogether, this proves the following theorem.
\begin{theorem}
There exists a monotone PTAS for machine covering on a constant
number of related machines, which runs in time linear in the number
of jobs.
\end{theorem}

\subsection{Amounts of jobs}
\label{sec:scale}

We are given a fixed (constant) number of machines $m$ of speeds
$s_1\geq\ldots\geq s_m$. (Since our PTAS will turn out to be truthful,
we may assume that we know the real speeds and can sort by them.)
Without loss of generality, we assume
that $s_1=1$. Note that the total size of all jobs may be
arbitrarily large. 
Let $n_0$ be the number of jobs of size strictly larger than $\OPT$,
the optimal value of the cover, in the input.
We begin by proving some auxiliary claims regarding $n_0$.
\begin{claim}
\label{cl:large}
$n_0\leq m-1$.
\end{claim}
\begin{proof}
Assume by contradiction that there are at least $m$ jobs that are all
larger than size $\OPT$. Assigning one job per machine, we get a
load larger than $\OPT$ on all machines (since all speeds are at most
1), which is absurd.
\end{proof}
%
%
\begin{claim}
\label{thesum}
The sum of sizes of all jobs that have size of at most $\OPT$ is 
at most $2\OPT(m-n_0-1)+\OPT$.
\end{claim}
\begin{proof}
Consider all jobs of size at most $\OPT$. Assume by contradiction
that the total size of these jobs is at least $2\OPT(m-n_0-1)+\OPT$. 
Let $A$ be an arbitrary set of jobs that some optimal algorithm puts
on some least loaded machine $j\in{1,\dots,m}$, and let $B$ be all
other jobs of size at most $\OPT$. By assumption, the total size of
the jobs in $B$ is more than $2\OPT(m-n_0-1)$. 
Since each job in $B$ has size at most $\OPT$, 
it is possible to partition these jobs
into sets, so that the total size of the first $m-n_0-1$ sets is in 
$(\OPT,2\OPT]$, and all remaining jobs are assigned to a set $C$
(which must be nonempty). This can for instance be done by sorting the
jobs in $B$ in order of decreasing size.
Assign each of the first $m-n_0-1$ sets to its own machine.
Assign the $n_0$ job larger than $\OPT$ to $n_0$ machines, one per
machine. Assign $A$ and $C$ to the remaining empty machine.
Since $C$ has nonzero size, we find an assignment with cover 
better than $\OPT$, a contradiction.
\end{proof}

\subsection{Finding a monotone \OPT}
\label{sec:monotone}

Let $\eps>0$ be a given constant. Without loss
of generality we assume $\eps<1.$
The algorithm in the next sections modifies the input so that we
end up with a constant number of jobs (at most
$4(m+2m^2/\eps^2$)).
The reason is that for this input, it is possible to enumerate
all possible job assignments in constant time (there are at most
$m^{4m+8m^2/\eps^2}$ different assignments). Before enumeration,
we define a fixed ordering on the machines. This ordering does
not need to depend on the speeds, and does not change even if
machine speeds are modified. Among all possible job assignments, we take the
optimal assignment which is lexicographically smallest among all
optimal assignment (using the fixed ordering). The usage of a
fixed ordering to obtain a monotone optimal algorithm was already
used for the makespan scheduling problem~\cite{ArcTar01}.

We show that this gives a monotone algorithm. Suppose machine $i$
claims to be faster, but it is not the bottleneck, then nothing
changes. The previous assignment is still optimal. A hypothetical
lexicographically smaller optimal assignment with the new speed 
would also reach a cover of the old optimal value with the old 
speed, because the old speed was lower, a contradiction.

If machine $i$ is the bottleneck (it is covered exactly to optimal
height), then $i$ will only get more work. This follows
because there are two options:

1. The algorithm concludes that the original assignment is still
the best (though with a smaller cover $C'$ than before), then the
amount allocated to $i$ remains unchanged.

2. The algorithm concludes that another assignment is now better,
then $i$ clearly gets more work (to reach a load above $C'$,
which is what $i$ has with the old amount of work and the old, slower speed).

\subsection{Reducing the number of jobs}
\label{sec:reduce}

We construct an input for which we can find an optimal job assignment
which is the smallest assignment lexicographically, and thus
monotone. We build it in a way that the value of an optimal
assignment for the adapted input is within a multiplicative factor
of $1-3\eps$ from the value of an optimal assignment for the
original input. This is done by reducing the number of jobs of
size no larger than $\OPT$ to a constant number (dependent on $m$ and
$\eps$), using a method which is oblivious of the machine speeds.

Let $\Delta={2m^2}/{\eps^2}+m$.
If the input consists of at most $\Delta$ jobs, then we are done. 
Otherwise, we keep the $\Delta$ largest such 
jobs as they are. This set is denoted by $J_L$. 
Let $J_S$ be the rest of the jobs.

Let $A$ be the total size of the jobs in $J_S$. Let $a$ be the size 
of the largest job in $J_S$. 

If $A\leq 3a\Delta$, we combine jobs greedily to create mega-jobs
of size in the interval $[a,3a]$. One mega-job is created by
combining jobs until the total size reaches at least $a$, this size
does not exceed $2\cdot a$. If we are left with a remainder of
size less than $a$, it is combined into a previously created job.
The resulting number of mega-jobs created from $J_S$ is at most
$3\Delta$.

Otherwise, we apply a ``List Scheduling'' algorithm with as input
the jobs in $J_S$ and $\Delta$ identical machines. These machines 
are only used to combine the jobs of $J_s$ into $\Delta$ mega-jobs
and should not be confused with the actual ($m$) machines in the
input. 

List Scheduling (LS) works by assigning the jobs one by one 
(in some order) to
machines, each job is assigned to the machine with minimum load
(at the moment the job is assigned).
LS thus creates $\Delta$ sets of jobs and the maximum difference
in size between two sets is at most $a$~\cite{Graham66}. The jobs in
each set are now combined into a mega-job. Thus we get $\Delta$ mega-jobs
with sizes in the interval $[\frac{A}{\Delta}-a,\frac{A}{\Delta}+a]$.
Since $\frac{A}{\Delta}\geq 3a$, we get that the ratio between the size
of two such mega-jobs is no larger than $2$.

In all three cases we get a constant number of jobs and mega-jobs.

\subsection{The optimal value of the modified instance}
\label{sec:optimal}

If no mega-jobs were created then clearly we consider all
possible job assignments and achieve an optimal one for the original
problem.
Consider therefore the two cases where we applied the jobs merging
procedure.
Note that since the total size of all jobs of size
at most $\OPT$  is at most $2m\OPT$ by Claim \ref{thesum}, 
and given the amount of jobs in
$J_L$ (and using Claim \ref{cl:large}), we have $a\leq \eps^2\OPT/m$.

First assume $A\leq 3a\Delta$.
We use the following notations. $\OPT'$ is the value of an optimal 
assignment using the modified jobs. $\OPT''$ is the value of an optimal
assignment using the modified jobs and only machines of speed at least
$2a/(\eps\OPT)$ (called fast, whereas all other machines are
called slow). Thus for $\OPT''$ we assume that the slow machines are
simply not present. Clearly we have $\OPT''\geq \OPT'$ and $\OPT\geq
\OPT'$.

We show that $\OPT'' \geq (1-2\eps) \OPT$. Given an optimal
assignment for the original instance, remove all
jobs assigned to slow machines. Remove all jobs that belong to
$J_s$ (which are of size at most $a$) that are assigned to fast
machines, and replace them greedily by mega-jobs. The mega-jobs
are assigned until that total size of allocated mega-jobs is just
about to exceed the total size of jobs of $J_s$ that were
assigned to this machine. Since all mega-jobs are of size at most
$4a$, and each fast machine has load of at least $\OPT$ and thus a
total size of assigned jobs of at least $2a/\eps$
(since it is fast), the
loss is at most of $2\eps$ of the total load. The rest of the
jobs (jobs of $J_L$ removed from slow machines, and remaining
mega-jobs) are assigned arbitrarily.

We next show how to convert an assignment with value $\OPT''$ 
(ignoring the slow machines) into an assignment which uses all
machines. Since there are at least $\Delta$ jobs of size at least
$a$ (the jobs of $J_L$), and these jobs are spread over at most
$m$ machines, at least one machine has at least ${\Delta}/{m}$
such jobs. From this machine, remove at most
$2m/\eps$ jobs of size at least $a$ (the smallest
ones among those that are large enough), and assign
$2/\eps$ jobs to each machine that does not
participate in the assignment of $\OPT''$. The resulting load of each
such machine (taking the speed into account) has a load of at
least $\OPT$ since it is slow: we have
$\frac{2}{\eps}\cdot a/(\frac{2a}{\eps\OPT})=\OPT$. 
The loss of the fast machine where jobs were
removed is at most a factor of $\eps$ of its original load.
Therefore we get that in the new job assignment
each machine is either loaded by at least
$\OPT$ or by at least $(1-{\eps})\OPT''$. Thus
$\OPT'\geq \min \{\OPT, (1-{\eps})\OPT'' \}$.
Since $\OPT''\geq (1-2\eps) \OPT$, this proves that $\OPT'\geq
(1-3\eps) \OPT$.

The second case is completely analogous, except that in this
case we call machines with speed at least 
$\left(\frac{A}{\Delta}-a\right)/(\eps\OPT)$ fast.
Thus each fast machine has total size of assigned jobs of at 
least $\left(\frac{A}{\Delta}-a\right)/\eps$.
We define fast in this way because in this case, the mega-jobs 
have size in the interval $[\frac{A}{\Delta}-a,\frac{A}{\Delta}+a]$.
When we replace jobs by mega-jobs, 
such a machine then loses at most $2\eps$ of its original load.
When we convert the assignment of $\OPT''$, we use that mega-jobs
have size at least ${\frac{A}{\Delta}-a}$, and there are $\Delta$
of them, so we can now transfer ${2m}/{\eps}$ of them to 
slow machines and get  the same conclusions as before.

\subsection{Running time}
\label{sec:time}

We reduce the number of jobs to a constant. Note in the reduction
in Section \ref{sec:reduce}, we are only interested in
identifying the $\Delta$ largest jobs. 
After this we merge all remaining jobs using a method based
on their total size. These things can be done in time linear in
$n$. Finally, once we have a constant number of jobs, we only
need constant time for the remainder of the algorithm. Thus our
algorithm has running time which is linear in the number of jobs
$n$.

\section{FPTAS for constant $m$}


In this section, we present a monotone fully polynomial-time
approximation scheme for constant $m$. This scheme uses
as a subroutine a non-monotone FPTAS which is described in
Section \ref{sec:nonfp}. We explain how this subroutine
can be used to create a monotone FPTAS in Section \ref{sec:monfp}.

In the current problem, it can happen that some jobs are superfluous:
if they are removed, the optimal cover that may be reached remains
unchanged. Even though these jobs are superfluous, 
we need to take special care of these jobs to make sure that our
FPTAS is monotone.
In particular, we need to make sure that these superfluous jobs
are always assigned in the same way, and not to very slow machines.
We therefore need to modify the FPTAS mechanism from~\cite{AnAzSo05}
because we cannot simply use any ``black box'' algorithm as was possible in~\cite{AnAzSo05}.

\subsection{An FPTAS which is not monotone}
\label{sec:nonfp}

Choose $\eps$ so that $1/\eps$ is
an integer. We may assume that $n\geq m$, otherwise $\OPT=0$ and we
assign all jobs to machine 1. In the
proof of Lemma \ref{SNCmono} we show that this assignment is
monotone.

We give an algorithm which finds the optimal cover up to a factor
of $1-2\eps$. We can again use an algorithm which is an
$m$-approximation~\cite{AzEp97}, therefore we can assume we can
find $\OPT$ within a factor of $m$. We scale the problem instance
such that our algorithm returns a cover of size 1. Then we know
that $\OPT \in [1,m]$. We are now going to look for the highest
value of the form $j\cdot\eps$ ($j=1/\eps,1/\eps+1,\dots,m/\eps$)
such that we can find an assignment which
is of value at least $(1-\eps)j\eps$. That is, we partition the
interval $[1,m]$ into many small intervals of length $\eps$. We
want to find out in which of these intervals $\OPT$ is, and find
an assignment which is at most one interval below it.
We can use binary search on $j$.

Given a value for $j$, we scale the input up by a factor of
$\frac{n}{j\eps^2}\geq \frac{m}{m\eps}\geq 1$. Now the target
value (the cover that we want to reach)
for a given value of $j$ is not $j\eps$ but $S=n/\eps$.
Sort the machines by speed. For machines with the same speed,
sort them according to some fixed external ordering.
For job $k$ and machine $i$, let $\ell_i^k=\lceil p_k/s_i\rceil$
($k=1,\dots,n;i=1,\dots,m$).

We use dynamic programming based on the numbers $\ell_i^k$. 
A \emph{load vector} of a given job assignment is an $m$-dimensional
vector of loads induced by the assignment.
Let $T(k,a)$ be a
value between 0 and $m$
for $k=1,\dots,n$ and an (integer!) load vector $a$.
$T(k,a)$ is the maximum number such that job $k$ is assigned to
machine $T(k,a)$ and a load vector of $a$ (or better) can be achieved with 
the jobs $1,\dots,k$. If the vector $a$ cannot be achieved, $T(k,a)=0$.

As soon as we find a value $k$ such that $T(k,S,\dots,S)>0$, we
can determine the assignment for the first $k$ jobs by going back
through the tuples (each time subtracting the last job from the
machine where it was assigned according to the value of the tuple).
If $k<n$, the last $n-k$ jobs are assigned to machine 1 (the fastest machine).

Now initialize $T(0,0)=m$ and $T(0,a)=0$ for any $a\geq0$.
For a load vector $a=(a_1,\dots,a_m)$,
$T(k,a)$ is computed from $T(k-1,a)$ by examining $m$ values
(each for a possible assignment of job $k$):
$$    T(k, a) = \max \left( 0, \left\{ i \in \{1,\dots,m\} \left|
a_i-\ell_i^k\geq0 \mbox{ and }
T(k-1, (a_{-i}, a_i - l^k_i)) > 0\right\} \right. \right)$$

Each value $T(k,\bar a)$ is set only once, i.e., if it is nonzero it is
not changed anymore.
The size of the table $T$ for one value of $k$ is $(S+1)^m$.
The $n$ tables are computed in total time
$nm(S+1)^m=O(m(n/\eps)^{m+1})$. Now note that the loss by
rounding is at most $n$ per machine. If we replace the rounded
jobs by the original (scaled) jobs, then the loss is at most 1
per job, and there are at most $n$ jobs on any machine.
Since the target value $S=n/\eps$,
we lose a factor of $1-\eps$ with regard to $S$.

In summary, we find the highest value of $j$ such that all
machines can be covered to $j\eps$ using jobs that are rounded.
The difference between the cover that we find and the cover of
the actual unrounded jobs is at most a factor of $\eps$ of this
cover, thus the actual cover found is at least $(1-\eps)j\eps$.
On the other hand, a cover of $(j+1)\eps$ cannot be
reached (not even with rounded jobs), so $\OPT<j\eps+\eps$. This
implies that our cover is at least $(1-\eps)(\OPT-\eps)\geq
(1-2\eps)\OPT$ since $\OPT\geq1$.

\subsection{A monotone FPTAS-mechanism}
\label{sec:monfp}

Our FPTAS mechanism is displayed in Figure \ref{fig:fptas}. 
As mentioned above, it is a variation on the FPTAS-mechanism described in~\cite{AnAzSo05}.
Their mechanism makes only one
direct reference to the actual goal function (makespan in their case)
and relies on a black box algorithm to find good assignments.
We made the following changes:
\begin{itemize}
\item Where the mechanism from~\cite{AnAzSo05} uses their black box
algorithm, we use instead the subroutine described in Section \ref{sec:nonfp}.
\item We need a different value for $\ell$, which denotes the second highest
power of $1+\eps$ that is considered as a valid bid.
We explain below how to find this value.
\item In the last step (testing all the sorted assignments),
we do not return the assignment with the
minimal makespan but instead the assignment with the maximal cover.
\end{itemize}

\begin{figure}
\begin{center}
\begin{fbox}{
\begin{minipage}{0.97\textwidth}
Input:
$n$ jobs in order of non-decreasing sizes,
a bid vector $b=(b_1,\dots,b_m)$, a parameter $\eps$ and a subroutine,
which is the FPTAS from Section \ref{sec:nonfp}.

\begin{enumerate}
\item Construct a new bid vector $d=(d_1,\dots,d_m)$ by rounding up
each bid to the closest value of $(1+\eps)^i$,
normalizing the bids such that the lowest bid is 1, and replacing
each bid larger than $(1+\eps)^{\ell+1}$ by $(1+\eps)^{\ell+1}$.
\item Enumerate over all possible vectors $d'=((1+\eps)^{i_1},\dots,
 (1+\eps)^{i_m})$, where $i_j\in\{0,\dots,\ell+1\}$. For each vector,
apply the subroutine and sort the output assignment such that
the $i$th fastest machine in $d'$ will get the $i$th largest amount of
work. 
\item Test all the sorted assignments on $d$, and return the one with
the maximal cover. In case of a tie, choose the assignment with the
lexicographically maximum assignment (where the machines are ordered
according to some external machine-id).
\end{enumerate}
\end{minipage}
}
\end{fbox}
\end{center}
\caption{\label{fig:fptas}A monotone FPTAS-mechanism}
\end{figure}

As specified in~\cite{AnAzSo05},
we will normalize the bids such that the lowest bid (highest speed) is 1.
Assuming the bids are truthful, i.e.~$b_j=1/s_j$ for
$j=1,\dots,m$, a very simple upper bound for the optimal cover is then
$U=\sum_{i=1}^n p_i$, the total size of all the jobs.
(Placing all the jobs on the fastest machine gives load
$U$ on that machine, and it is clear that the fastest
machine cannot get more load than this.)

Consider a slower machine $j$. Suppose $b_j\geq
U/p_n$. Then the load of this machine if it receives
only job $n$ is at least $U\geq\OPT$. This means that for our algorithm,
it is irrelevant what the exact value of $b_j$ is in this case,
because already for $b_j=U/p_n$ an optimal cover is certainly
reached by placing a single arbitrary job on machine $j$.
We can therefore change any bid which is higher than $U/p_n$
to $U/p_n$.

Since the mechanism normalizes and
rounds bids to powers of $1+\eps$, we can now define
$$\ell=\left\lceil\log_{1+\eps} \frac U{p_n}\right\rceil =
 \left\lceil\log_{1+\eps} \frac{\sum_{i=1}^n p_i}{p_n}\right\rceil.
$$
Plugging this in in the mechanism from~\cite{AnAzSo05},
this gives us a fully polynomial-time approximation scheme
for the machine covering problem, since $\ell$ is still (weakly) polynomial
in the size of the input. 
We prove in the appendix that it is monotone,
using a proof similar to Andelman et al.~\cite{AnAzSo05}.

\begin{theorem}
This FPTAS-mechanism is monotone.
\end{theorem}
\begin{proof}
We follow the proof of Andelman et al.~\cite{AnAzSo05}.
We need to adapt this proof to our goal function.
Suppose that machine $j$ increases its bid.
First of all, if the increase is so small that the vector $d'$
remains unchanged, the subroutine will give the same
output, and in step 3 we will also choose the same assignment.
Thus the load on $j$ does not change.

If $d_j>(1+\eps)^\ell$, the assignment found by our algorithm will
also not change when $j$ slows down: the vector $d'$
again remains the same and we can reason as in the first case.

Now suppose that $d_j\leq(1+\eps)^\ell$, and the speed of $j$
changes so that its rounded bid increases by a factor of $1+\eps$.
(For larger increases, we can apply this proof repeatedly.)
Suppose that $j$ is not the unique fastest machine.
We thus consider the case where a normalized rounded bid rises from
$d_j$ to $(1+\eps)d_j$, the assignment changes from $W$ to $W'$, and
we assume that the amount of work assigned to machine $j$ increases from
$w_j$ to $w_j'>w_j$. Denote the size of the cover of assignment $W$
on bid vector $d$ by $C$. There are two cases.

Suppose that the cover that our algorithm finds
increases as $j$ becomes slower.
So all machines have load strictly above $C$.
Consider the new assignment $W'$ on the old speeds.
All machines besides $j$ do not change their speeds and therefore
still have a load strictly above $C$.
Machine $j$ receives more work than in the old assignment $W$
and therefore also has a load strictly above $C$, since
it already had at least $C$ when it was faster.
This means that $W'$ gives a better cover than $W$ on the old speeds.
However, our algorithm would then have output $W'$ in the first
place, because it checks all these speed settings, a contradiction.

Now suppose that the cover that our algorithm finds
stays the same as $j$ becomes slower.
This means that $j$ is not the bottleneck machine (the unique least
loaded machine).
The old assignment $W$ clearly has a cover of $C$ also with the new speeds,
so our algorithm considers it. It would only output $W'$ if
$W'$ were lexicographically larger than $W$ and also had a cover of $C$
(or better). However, in that case $W'$ again would have been found
before already exactly as above, a contradiction.

Finally, suppose that $j$ is the unique fastest machine.
Due to normalization, $d_j$ remains 1, bids between $1+\eps$ and
$(1+\eps)^\ell$ decrease by one step, and bids equal to
$(1+\eps)^{\ell+1}$ can either decrease to $(1+\eps)^\ell$ or remain
unchanged.
We construct an alternative bid vector $\hat d$ as in~\cite{AnAzSo05}
where we replace all bids of $(1+\eps)^{\ell+1}$ in $d'$ with $(1+\eps)^\ell$.
This is the point where we use the fact that we check ``too many''
speed settings.

Every machine that bids $(1+\eps)^{\ell}$ or more needs to receive only at
least one arbitrary job to have sufficient load.
In such cases, our subroutine indeed puts only one job on such a machine,
because it finds the minimum amount of jobs $k$ to get to a certain cover
and puts all remaining jobs on the fastest machine.
Therefore, the cover that our algorithm finds
for $\hat d$ will be the same as that for $d'$,
and it will also give the same output assignment. This is also optimal
for $(1+\eps)\hat d$. The difference between $(1+\eps)\hat d$ and
$d$ is only that the bid $d_j$ changes from 1 to $1+\eps$.
We can now argue as before: whether the cover that our algorithm finds
increases or not as $j$ becomes slower, a hypothetical new better
assignment for $\hat d(1+\eps)$ would also be better for $d$,
but in that case the algorithm would have found it before.
\end{proof}

\section{Approximation algorithm SNC for arbitrary values of $m$}

We present an efficient approximation algorithm for an arbitrary
number of machines $m$. Our algorithm uses Next Cover~\cite{AzEp97}
as a subroutine.
This semi-online algorithm is defined in Figure \ref{fig:nc}.
Azar and Epstein~\cite{AzEp97} showed that if the optimal cover
is known, Next Cover (NC) gives a 2-approximation. That is, for
the guess $G=\OPT/2$ it will succeed.
\begin{figure}
\begin{center}
\begin{fbox}{
\begin{minipage}{0.97\textwidth}
Input: guess value $G$, $m$ machines in a fixed order of non-increasing speeds,
$n$ jobs in order of non-increasing sizes.

For every machine in the fixed order, starting from machine 1,
allocate jobs to the machine according to the sorted order of jobs until the
load is at least $G$.

If no jobs are left and not all machines reached a load level of $G$, report
failure. If all machines reached a load of $G$, allocate remaining jobs (if any)
to machine $m$, and report success.
\end{minipage}
}
\end{fbox}
\end{center}
\caption{\label{fig:nc}Algorithm Next Cover (NC)}
\end{figure}
NC also has the following property, which we will use later.
\begin{lemma}
\label{lem:nc} Suppose NC succeeds with guess $G$ but fails with
guess $G+\eps$, where $\eps\leq\frac13 G$. Then in the
assignment for guess $G$, the work on machine $m$ is less than
$mw+\eps$, where $w\geq G$ is the minimum work on any
machine.
\end{lemma}
\begin{proof}
Consider machine $m$. Suppose its work is at least $mw+\eps$,
where $\eps\leq\frac{G}{3}\leq\frac{w}3$.

Suppose $m$ is odd. We create a new assignment as follows. Place
the jobs on machines $i,i+1$ on machine $(i+1)/2$ for
$i=1,3,5,\dots,m-2$. Cut the work on machine $m$ into $(m+1)/2$
pieces (without cutting any jobs) that all have size at least
$w+\eps$ and at most $2w$. Put these on the last $(m+1)/2$ machines.

The proof that it is possible to cut the pieces in this way is
analogous to that for set $B$ in the proof of Claim \ref{thesum}.
The last piece then has
size at least $mw+\eps - \frac{m-1}2\cdot2w = w+\eps$. This
means that NC succeeds with guess $w+\eps\geq G+\eps$, a
contradiction.

Now suppose $m$ is even. This time we create a new assignment by
placing the jobs on machines $i,i+1$ on machine $(i+1)/2$ for
$i=1,3,5,\dots,m-3$. \Xomit{On machine $m/2$, we put the jobs of
machine $m-1$ plus the first $k$ jobs on machine $m$, where $k$
is the maximum integer such that the total work of these $k$ jobs
is at most $w$ (we have $k\geq1$). The work on machine $m/2$
is then at least $w+\eps$, which follows from an easy case
distinction on $k$. Cut the remaining work on machine $m$ into
$m/2$ pieces that all have size at least $w+\eps$. This is
possible, because after cutting $m/2-1$ times, we have cut off at
most $w + (m/2-1)2w = w(1+m-2) = w(m-1)$, (the first
$w$ comes from cutting off the first $k$ jobs), leaving at
least $w+\eps$ for the last machine. This gives a
contradiction as above.} Note that machine $m-1$ already has jobs
no larger than $w$. That is true since some machine $i$ among
$1,\ldots,m-1$ has received work of exactly $w$, and all jobs
assigned to machines $i,\ldots,m$ are no larger than $w$. We
can consider the total work of the last two machines. This load
is at least $(m+1)w+\eps$ and as shown before, it can be split
into $\frac{m+2}{2}=\frac m2+1$ parts of size at least
$w+\eps$ each. The parts can be assigned in the appropriate
order to machines $\frac m2,\ldots,m$.
\end{proof}

Our algorithm Sorted Next Cover (SNC) works as follows.
A first step is to derive a lower bound and an upper bound on the
largest value which can be achieved for the input and $m$
identical machines. To find these bounds, we can apply LPT
(Longest processing Time), which assigns the sorted (in
non-increasing order) list of jobs to identical machines one by
one. Each job is assigned to the machine where the load after
this assignment is minimal.
It was shown in \cite{DFL82,CKW92} that the approximation
ratio of LPT is $\frac{4m-2}{3m-1}<\frac 43$. Thus we define $A$
to be the value of the output assignment of LPT. We also define
$L=\frac A2$ and $U=\frac 43 A$. We have that $A$ and $U$ are
clear lower an upper bounds on the optimal cover on identical
machines. Since NC always succeeds to achieve half of an optimal
cover, it will succeed with the value $G=L$. Since a cover of $U$
is impossible, the algorithm cannot succeed with the value $G=U$.
Throughout the algorithm, the values $L$ and $U$ are such that
$L$ is a value on which NC succeeds whereas $U$ is a failure
value. We perform a geometrical binary search. It is possible to
prove using induction that if NC succeeds to cover all machines
with a guess value $G$, then it succeeds to cover all machines
using a smaller guess value $G'<G$. The induction is on the
number of machines and the claim is that in order to achieve a
cover of $G'$ on the first $i$ machines, NC uses the same subset
or a smaller subset used to achieve $G$.

The algorithm has a parameter $\eps\in(0,1/2)$ that we can set
arbitrarily. See Figure \ref{fig:snc}. Since the ratio between
$U$ and $L$ is initially constant, it can be seen that the
algorithm completes in at most $O(\frac {1}{\log (1+\eps/2)})$
steps. The overall running time is
$O(n(\log n + 1/\log (1+\eps/2))$ due to the sorting.
Note that Steps 2 and 6 are only executed once.

\begin{figure}
\begin{center}
\begin{fbox}{
\begin{minipage}{0.97\textwidth}
Input: parameter $\eps\in(0,1/2)$, sorted set of jobs
($p_1\geq\dots\geq p_n$), sorted machine bids ($b_1\leq\dots\leq
b_m$).
\begin{enumerate}
\setlength{\itemsep}{0pt}
\item If there are less than $m$ jobs, assign them to machine 1 (the machine of speed $s_1$),
output 0 and halt.
\item Scale the jobs so that $\sum_{i=1}^n p_j=1$.
Run LPT on identical machines and denote the value of the output by $A$.
Set $L=\frac {A}{2}$ and $U=\frac 43 A$. 
\item Apply Next Cover on identical machines with the guess $G=\sqrt{U \cdot L}$.
\item If Next Cover reports success, set $L=G$,
else set $U=G$.
\item If $U-L>\frac\eps2 L$, go to step 3, else continue with step 6.
\item Apply Next Cover on identical machines with the value $L$.
Next Cover partitions the jobs in $m$ subsets, each of total size
of jobs at least $L$.
Sort the subsets in non-increasing order and allocate
them to the machines in non-increasing order of speed according
to the bids.
\end{enumerate}
\end{minipage}
}
\end{fbox}
\end{center}
\caption{\label{fig:snc}Algorithm Sorted Next Cover (SNC)}
\end{figure}

\begin{lemma}
\label{SNCmono} SNC is monotone.
\end{lemma}

\begin{proof}
The subsets constructed in step 3 and 6 do not depend on the
speeds of the machines. If a machine claims it is faster than it
really is, the only effect is that it may get a larger subset.
Similar if it is slower.

If the algorithm halts in step 1, then we again have a situation
that jobs are partitioned into sets, and the sets are assigned in
a sorted way. This is actually the output that steps 2--6 would
produce if SNC was run with a guess value $0$.
\end{proof}

\begin{theorem}
\label{th:snc}
For any $0<\eps<1$, SNC maintains an approximation ratio of $\min(m,(2+\eps)s_1/s_m)$.
\end{theorem}
\begin{proof}
We start with the second term in the minimum.
The load that SNC has on machine $i$ is at least $L/s_i$,
and Next Cover cannot find a cover above  $U\leq (1+\eps/2)L$
on identical machines.
So the optimal cover on identical machines of speed 1 is at most
$2(1+\eps/2)L= (2+\eps)L$.
Thus the optimal cover on machines of speed $s_m$ is at most
$(2+\eps)L/s_m$, and the optimal cover on the actual machines
can only be lower since $s_m$ is the smallest speed.
We thus find a ratio of at most
$((2+\eps)L/s_m)/(L/s_i) = (2+\eps)s_i/s_m\leq (2+\eps)s_1/s_m.$

We prove the upper bound of $m$ using induction.

\emph{Base case:} On one machine, SNC has an approximation ratio of 1.

\emph{Induction hypothesis:} On $m-1$ machines, SNC has an
approximation ratio of at most $m-1$.

\emph{Induction step:}
%
Recall that the jobs are scaled so that their total size is 1.
Suppose each machine $j$ has work at least $1/(jm)$ ($j=1,\dots,m$).
Then the load on machine $j$ is at least $1/(j m s_j)$.
However, the optimal cover is at most
$1/(s_1 + s_2 + ... + s_m) \leq 1/(j s_j + (m-j) s_m) \leq 1/(j s_j)$.
Thus SNC maintains an approximation ratio of at most $m$ in this case.

Suppose there exists a machine $i$ in the assignment of SNC
with work less than $1/(im)$.
Consider the earliest (fastest) such machine $i$. Due to the resorting
we have that the work on machines $i,\dots,m$ is less than $1/(im)$.
So the total work there is less than $(m-i+1)/(im)$.
The work on the first $i-1$ machines is then at least
$1 - (m-i+1)/(im) = (im-m+i-1)/(im) = (i-1)(m+1)/(im)$ and
the work on machine 1 is at least $(m+1)/(im)$.
This is more than $m+1$ times the work on machine $i$.

We show that in this case there must exist a very large job, which
is assigned to a machine by itself.
Let $L'$ and $U'$ be the final values of $L$ and $U$ in the
algorithm. Let $w$ be the minimum work assigned to any machine for
the guess value $L'$. Since SNC gives machine $i$ work less than
$1/(im)$, we have $w<1/(im)$. We have $U'-L'\leq \frac\eps2 L'$.
SNC succeeds with $L'$ and fails with $U'$ and thus, since $\eps
\leq \frac 12$ and by Lemma \ref{lem:nc}, machine $m$ receives at
most $mw+\frac\eps2L' \leq mw+\frac14L'\leq (m+\frac 14)w \leq
(m+\frac 14)/(im)$ running NC with the guess value $L'$. Moreover,
NC stops loading any other machine in step 6 as soon as it covers
$L'$.

We conclude that the only way that any machine can get work more than $(m+1)L'$
is if it gets a single large job.
This means that in particular the first (largest) job has size
$p_1>(m+1)w\geq 3w\geq
3L' $. SNC assigns this job to its first machine, and the remaining
work on the other machines.

To complete the induction step, compare the execution of SNC to
the execution of SNC with as input the $m-1$ slowest machines and
the $n-1$ smallest jobs. Denote the first SNC by SNC${}_m$ and the
second by SNC${}_{m-1}$.
We first show that SNC${}_{m-1}$ 
fails on $U'$. Since
$U'\leq (1+\frac\eps2)w <2w$, then SNC${}_m$ 
assigns only $p_1$ to machine 1, and thus SNC${}_{m-1}$ 
executes exactly the same on the other machines. Since machine 1
is covered, SNC${}_m$  
fails on some later machine,
and then this also happens to SNC${}_{m-1}$.
Therefore, SNC${}_{m-1}$ 
cannot succeed with $U'$ or any larger value.
A similar reasoning shows that  SNC${}_{m-1}$ succeeds with any guess that
is at most $L'$.
%
Finally, $L'$ is at least the starting guess $A/2$. So
$p_1>3L'\geq\frac32 A$ implies that LPT also puts only the first job
on the first machine, since its approximation ratio is better than $4/3$.
Therefore, LPT gives the same guess value $A$ for the original input on $m$
machines as it would for the $n-1$ smallest jobs on $m-1$ machines.
%
This means that SNC${}_m$ and SNC${}_{m-1}$ 
maintain the same values $U$ and $L$
throughout the execution, and then we can apply the
induction hypothesis.
\Xomit{ If $p_j\geq1/m$, then each time that Step 3 is executed,
both algorithms produce the same assignment, except that the SNC
with $m$ machines has an extra machine with only job $j$. This
means that both algorithms maintain the same guess value
throughout, and eventually output the same assignment in Step 6
(ignoring the fastest machine).

Suppose $p_j<1/m$. We know that the final guess $L$ of SNC is
less than $p_j/3$. Thus as long as the guess value is at least
$p_j$, both algorithms (SNC with and without the fastest machine
and largest job) must fail in Step 3, and halve their guess and
their $\eps$. As soon as the guess value drops below $p_j$, SNC
always puts only the largest job on the first machine, and the
remaining assignment is identical to the assignment of SNC without
that job and that machine.

This means that the two algorithms maintain the same values $U,L$
and $\eps$ throughout the execution, and then we can apply the
induction hypothesis.}
\end{proof}

Next we show that the simple algorithm Round
Robin has an approximation guarantee of $m$, so this algorithm
can also be used in case the speed ratio is large.
It should be noted that if we find an algorithm with a better
guarantee than $m$, we cannot simply run both it and SNC and
take the best assignment to get a better overall guarantee. The
reason that this does not work is that this approach does not
need to be monotone, even if this hypothetical new algorithm
is monotone: we do not know what happens at the point where we
switch from one algorithm to the other.

\paragraph{Round Robin}
Sort the machines and jobs by speed, so that the first machine
has the largest speed and the first job has the largest size. The
Round Robin algorithm assigns jobs of indices $i+mk$ (in the
sorted list) to machine $i$ (in the sorted list) for $k\geq 0$
until it runs out of jobs. Comparing two successive machines, we
see that the $j$th job on machine $i+1$ is never larger than the
$j$th job on machine $i$ (and may not even exist at all in case
we ran out of jobs). Thus the work is monotonically decreasing.
Moreover, the job sets that are constructed are independent of
the speed, and the only effect of e.g. bidding a higher speed is
to possibly get a larger set of jobs. Thus this algorithm is
monotone.

\begin{claim}
The approximation ratio of Round Robin is exactly $m$.
\end{claim}
\begin{proof}
It is easy to see that the ratio cannot be better than $m$.
Consider $m$ identical machines, $m-1$ jobs of size 1 and $m$
jobs of size $1/m$. Round Robin places only one job of size $1/m$
on the last machine and has a cover of $1/m$. By placing all the
small jobs on the last machine, it is possible to get a cover of
1.

Consider the first machine in the ordering. It gets at least a
fraction of $1/m$ of the total size of all jobs. Consider now
another machine, whose index in the ordering is $i$. We change
the sequence in the following way. Take the largest $i-1$ jobs
and enlarge them to size $\infty$. Clearly, \OPT can only
increase. Call these jobs ``huge''. Next, we claim that without
loss of generality, huge jobs are assigned to the first $i-1$
machines in the ordering by \OPT. Otherwise, do the following
process. For $j=1,...,i-1$, if machine $j$ has a huge job, do
nothing. Otherwise, remove a huge job from a machine $x$ in
$i,...,m$ (again, indices are in the sorted list), and put it on
machine $j$, put the jobs of machine $j$ on machine $x$. Since
$j$ is not slower than $x$, the cover does not get smaller. We
got an assignment $\OPT'\geq \OPT.$ Consider now the assignment the
algorithm creates. Consider only the jobs which are not huge, we
placed these jobs in a Round-Robin manner, starting from machine
$i$. Therefore, machine $i$ received at least an $1/m$ fraction
of these jobs (with respect to total size). On $\OPT'$, machine
$i$ does not have huge jobs, thus it can have at most $m$ times
as much work as in our assignment. Thus we have a cover of at least
$\OPT'/m\geq \OPT/m$.
\end{proof}

\section{Algorithms for small numbers of machines}
\label{sec:small}

We next consider the case of two machines. Even though previous
sections give algorithms for this case with approximation ratio
arbitrarily close to $1$, we are still interested in studying the
performance of SNC for this case. The main reason for this is that
we hoped to get ideas on how to find
algorithms with good approximation ratios for $m>2$ machines
that are more efficient
than our approximation schemes. However, as we show below,
several obvious adaptations of SNC are not monotone, and it
seems we will need more sophisticated algorithms for $m>2$.

A first observation is that
there are only $n-1$ possible partitions of the jobs into two
sets (since we keep the jobs in sorted order),
and thus there is no need to perform binary search.
Let $S_i=(L_i=\{1,\ldots,i\},R_i=\{{i+1},\ldots,n\})$ be a
partition of the sorted list of jobs ($p_1\geq p_2\dots\geq
p_n$). Clearly, to have a finite approximation ratio we only need to
consider $S_i$ for $i=1,\ldots,n-1$. For a given partition $S_i$,
let $\sigma_1(i)=\sum_{j=1}^i p_j$ and
$\sigma_2(i)=\sum_{j=i+1}^n p_j$.

SNC is defined for two machines as follows. See Figure \ref{fig:snc2}.
\begin{figure}
\begin{center}
\begin{fbox}{
\begin{minipage}{0.97\textwidth}
Input: sorted set of jobs
($p_1\geq\dots\geq p_n)$, sorted machine bids ($b_1\leq b_2$)

Find $i$ such that
$\min\{\sigma_1(i),\sigma_2(i)\}$ is maximal.
If $\sigma_1(i) \geq \sigma_2(i)$, assign $L_i$ to the first (faster)
machine and $R_i$ to the second. Else, assign $L_i$ to the second
machine and $R_i$ to the first. 
\end{minipage}
}
\end{fbox}
\end{center}
\caption{\label{fig:snc2}Algorithm Sorted Next Cover (SNC) on two machines}
\end{figure}
 From Theorem \ref{th:snc}
it follows that SNC (which ignores the speeds) has an
approximation of at most $2$.
\Xomit{
An example for large speeds shows that the overall bound cannot
be improved. Consider machines with speeds $s\geq 2$ and $1$, and
the jobs $\frac s2,\frac s2,1$. SNC will create the sets $\{\frac
s2,1\}$, $\{\frac s2\}$, which gives a cover of $\frac 12+\frac
1s$ whereas $\OPT=1$, thus the approximation ratio tends to $2$.
}
We next consider the approximation ratio as a function of the speed
ratio $s\geq 1$. 

\begin{lemma}
\label{lem:snc2}
On two machines, SNC has an approximation ratio of
$\max\{\frac{3}{s+1},\frac{2s}{s+1}\}$.
\end{lemma}
\begin{proof}
Assume without loss of generality that the speeds are $s$ and $1$.
Since the total work is 1, we have $\OPT\leq \frac{1}{s+1}$.

Let $i$ be the index such that the partition chosen by SNC is
$S_i$. We have that the set of jobs which is assigned to $M_1$,
has the sum $\max\{\sigma_1(i),\sigma_2(i)\}\geq \frac{1}{2}$.
Thus if $M_1$ has a smaller load than $M_2$, this load is at least
$\frac{1}{2s}$ and we have an approximation ratio of at most
$\frac{\OPT}{1/(2s)}\leq \frac{2s}{s+1}$.

To give a lower bound on the load of $M_2$, consider first the
amount of jobs of size larger than $\frac{1}{3}$ in the input. If
no such jobs exist, let $j$ be the smallest index $1\leq j \leq
n-1$, such that $\sigma_1(j)\geq \frac 13$. Clearly $j$ exists
since $\sigma_1(n)= 1$. We would like to show that $\sigma_1(j) <
\frac 23$. If $\sigma_1(j) = \frac 13$ we are done, otherwise,
$j\geq 2$ since $p_1<\frac 13$. We have $\sigma_1(j-1) < \frac
13$ and thus $\sigma_1(j)= \sigma_1(j-1)+p_j < \frac 13+\frac
13=\frac 23$. Thus
\begin{equation}
\label{eq:sigma}
\min\{\sigma_1(i),\sigma_2(i)\}\geq
\min\{\sigma_1(j),\sigma_2(j)\}\geq \frac 13.
\end{equation}

Consider the case where there are two such jobs, thus $p_1\geq
p_2 >\frac 13$, or there is a single such job $p_1$ but $p_1\leq
\frac 23$,  we have $\sigma_1(1)>\frac 13$ and $\sigma_2(1)>\frac
13$ and thus again (\ref{eq:sigma}) holds.
Finally, in case $p_1>\frac 23$,
clearly $i=1$. We get that $\OPT\leq \sigma_2(1)$ and thus $M_2$
has (at least) optimal load.

Suppose $p_1\leq\frac 2 3$. Then
by (\ref{eq:sigma}) we have
$\sigma_2(i)\geq \frac 1 3$. 
This implies that if $M_2$ has load smaller than $M_1$, we have
an approximation ratio of at most $\frac{\OPT}{1/3}\leq
\frac{3}{s+1}$.

To show that the bound is tight, consider the following sorted
sequences. The first sequence consists of $\frac 12$ and the two
jobs $\frac{s-1}{2(s+1)}$ and $\frac{1}{s+1}$ if $s\geq 3$ (or
$\frac 12,\frac{1}{s+1},\frac{s-1}{2(s+1)}$ if $s<3$). An optimal
assignment assigns $\frac 1{s+1}$ to $M_2$ and the other two jobs
to $M_1$, thus $\OPT=\frac 1{s+1}$. However, SNC partitions the
input into two sets whose sizes are $\frac 1 2$, and so the
approximation ratio is $\frac{2s}{s+1}$.

The second sequence needs to be shown only for $s\leq \frac 32$.
We use the sorted sequence $\frac 13,\frac
13,\frac{2s-1}{3s+3},\frac{2-s}{3s+3}$ (this is a sorted sequence
for any $s\leq 2$). There are two possible best partitions, but
for both of them, the minimum work is on $M_2$ and is $\frac 13$.
However, an optimal assignment assigns one job of size $\frac 13$
and a job of size $\frac{2s-1}{3s+3}$ to $M_1$, and the other
jobs to $M_2$, getting a cover of $\frac{1}{s+1}$. We get an
approximation ratio of $\frac{3}{s+1}$.
\end{proof}

Below we prove that the fact that SNC ignores the
speeds is crucial for its monotonicity in the general case.
However, if $m=2$, we can
define an algorithm SSNC which takes the speeds into account and
is monotone as well.
SSNC is defined in Figure \ref{fig:ssnc}.

\begin{figure}
\begin{center}
\begin{fbox}{
\begin{minipage}{0.97\textwidth}
Input: sorted set of jobs
($p_1\geq\dots\geq p_n)$, sorted machine bids ($b_1\leq b_2$)

Let $r=b_2/b_1\geq 1$ be the speed ratio between the two machines
according to the bids. Find
$i$ such that $\min\{\frac{\sigma_1(i)}{r},\sigma_2(i)\}$ is
maximal. If $\sigma_1(i) \geq \sigma_2(i)$, assign $L_i$ to the
first (faster) machine and $R_i$ to the second. Else, assign
$L_i$ to the second machine and $R_i$ to the first.
\end{minipage}
}
\end{fbox}
\end{center}
\caption{\label{fig:ssnc}Algorithm Speed-aware Sorted Next Cover
(SSNC) on two machines}
\end{figure}

\begin{lemma}
Let $i$ indicate the partition that SSNC outputs for speed ratio $r$.
Then
\begin{equation}
\label{eq:head}
\frac{\sigma_1(i)}{r} \geq
\sigma_2(i)-p_{i+1}
\end{equation}
and
\begin{equation}
\label{eq:tail}
\sigma_1(i)- p_i  \leq r\sigma_2(i).
\end{equation}
\end{lemma}
\begin{proof}
Since $i$ was a best choice,
 $\min
\{\frac{\sigma_1(i)}{r},\sigma_2(i)\} \geq \min
\{\frac{\sigma_1(i)+p_{i+1}}{r},\sigma_2(i)-p_{i+1}\}$.
Since $p_{i+1}>0$, this implies $\min
\{\frac{\sigma_1(i)+p_{i+1}}{r},\sigma_2(i)-p_{i+1}\}=\sigma_2(i)-p_{i+1}$.
Filling this in in the inequality proves (\ref{eq:head}).

Similarly, we have $\min
\{\frac{\sigma_1(i)}{r},\sigma_2(i)\} \geq \min
\{\frac{\sigma_1(i)-p_{i}}{r},\sigma_2(i)+p_{i}\}$ which implies
$\min \{\frac{\sigma_1(i)-p_{i}}{r},\sigma_2(i)+p_{i}\}
=\frac{\sigma_1(i)-p_{i}}{r}$, leading to (\ref{eq:tail}).
\end{proof}

\begin{theorem}
\label{ssnc_2_mach}
SSNC is monotone on two machines.
\end{theorem}
\begin{proof}
As a first step we show the following. Let $s_1\geq s_2$ and $q_1
\geq  q_2$ be two speed sets such that $r_s=\frac
{s_1}{s_2}>r_q=\frac{q_1}{q_2}$. Let $i_s$ and $i_q$ be the
partitions which SSNC outputs for $r_s$ and $r_q$ respectively.

We show the following: $\max \{\sigma_1(i_s),\sigma_2(i_s)\} \geq
\max \{\sigma_1(i_q),\sigma_2(i_q)\}$ and $\min
\{\sigma_1(i_s),\sigma_2(i_s)\} \leq \min
\{\sigma_1(i_q),\sigma_2(i_q)\}$. Since
$\sigma_1(i_s)+\sigma_2(i_s)=\sigma_1(i_q)+\sigma_2(i_q)$, it is
enough to show one of the two properties.
Clearly, if $i_s=i_q$ this holds, therefore we assume that
$i_s\neq i_q$. Furthermore, we show that in this case we have
$i_s>i_q$.

Assume that $i_s<i_q$. Then $\sigma_1(i_s)<\sigma_1(i_q)$ and $\sigma_2(i_s)>\sigma_2(i_q)$.
By definition of the algorithm we have
$\min \{\frac{\sigma_1(i_s)}{r_s},\sigma_2(i_s)\} \geq \min
\{\frac{\sigma_1(i_q)}{r_s},\sigma_2(i_q)\}$ and $\min
\{\frac{\sigma_1(i_s)}{r_q},\sigma_2(i_s)\} \leq \min
\{\frac{\sigma_1(i_q)}{r_q},\sigma_2(i_q)\}$.
To avoid contradiction, we must have $\min
\{\frac{\sigma_1(i_q)}{r_s},\sigma_2(i_q)\}=\sigma_2(i_q)$ and
$\min
\{\frac{\sigma_1(i_s)}{r_q},\sigma_2(i_s)\}=\frac{\sigma_1(i_s)}{r_q}$.
Filling this in in the inequalities gives
$\frac{\sigma_1(i_s)}{r_s} \geq\sigma_2(i_q)$ and
$\frac{\sigma_1(i_s)}{r_q}\leq \sigma_2(i_q)$. This implies
$r_q\geq r_s$, a contradiction.

We may conclude $\min \{\sigma_1(i_s),\sigma_2(i_s)\} \leq
\sigma_2(i_s)\leq \sigma_2(i_q)-p_{i_q+1}\leq \sigma_1(i_q)$,
where the last inequality follows from (\ref{eq:head}), and
$\sigma_2(i_s)<\sigma_2(i_q)$, thus $\min
\{\sigma_1(i_s),\sigma_2(i_s)\} \leq \min
\{\sigma_1(i_q),\sigma_2(i_q)\}$.

Suppose \emph{$M_2$ becomes slower}. Then the speed ratio between
the two machines becomes larger. $M_2$ is still the slower
machine and thus by the above, the amount of work it gets cannot
increase.

Now suppose \emph{$M_1$ becomes slower}. We may assume $M_1$
remains faster than $M_2$. Otherwise, we divide the slowing down
into three parts. The first part is where $M_1$ is still faster
than $M_2$. In the middle part, the speeds do not change, but we
change the order of the machines. Clearly, at this point the work
on $M_1$ does not increase. Finally $M_1$ slows down further, but
now we can use the analysis from above because it is like $M_2$
getting slower.

Thus $M_1$ is still faster than $M_2$ but the speed ratio
decreases. By the statement above, we get that the amount of work
that $M_1$ gets cannot increase.
\end{proof}

\begin{theorem}
\label{th:ssnc2}
On two machines, SSNC has an approximation ratio of at most
$\min\{1+\frac{s}{s+1},1+\frac 1 s\}$.
\end{theorem}
\begin{proof}
Consider an optimal
assignment, and let $\mu$ the sum of jobs assigned to $M_1$ by this
assignment. Since the total work is 1,
the sum of jobs assigned to $M_2$ is $1-\mu$
and $\OPT=\min\{\frac{\mu}{s},1-\mu\}\leq \frac{1}{s+1}$.

\vspace{10pt}
\noindent Consider first the case $s\geq \phi$. We claim that there exists an
integer $1\leq i'\leq n-1$ such that
\begin{equation}
\label{eq:iprime}
\frac{s\cdot
\OPT}{s+1}\leq\sigma_2(i') \leq \frac{s\cdot \OPT}{s+1}+(1-\mu).
\end{equation}
Consider the smallest index $j$ of an item $p_j\leq 1-\mu$.
Clearly, $j\leq n-1$ since the optimal assignment we consider
assigns an amount of exactly $1-\mu$ to $M_2$, and moreover, by
the same reasoning, $\sigma_2(j)\geq 1-\mu$. If $j$ satisfies the
condition (\ref{eq:iprime}), we define $i'=j$ and we are
done.
If  $\sigma_2(j) < \frac{s\cdot \OPT}{s+1}$ we find
$\OPT=\min\{\frac{\mu}{s},1-\mu\}\leq 1-\mu
\leq \sigma_2(j) < \frac{s\cdot \OPT}{s+1} < \OPT$, a contradiction.

We are left with the case
$\sigma_2(j) > \frac{s\cdot \OPT}{s+1}+(1-\mu)$.
Let $j'$ such that $j<j'\leq n$ be the smallest index for which
$\sigma_2(j') < \frac{s\cdot \OPT }{s+1}$ (note that we allow
$j'=n$ which does not give a valid partition). Since $j'>
j$, we have $p_{j'}\leq 1-\mu$ and thus $\sigma_2(j'-1) =
\sigma_2(j')+p_{j'} < \frac{s\cdot \OPT }{s+1}+1-\mu$. In this
case define $i'=j'-1\leq n-1$.

We next show that $\sigma_1(i')\geq \frac{s^2\cdot \OPT}{s+1}$, and
later show that this implies the approximation ratio. Note that
by the definition of $i'$ we have $\sigma_1(i') \geq
\mu-\frac{s\cdot \OPT}{s+1}$.
There are two cases.
If $\mu \geq \frac{s}{s+1}$, we have $\OPT=1-\mu\leq
\frac{1}{s+1}$. We then find $\sigma_1(i') \geq 1-\OPT-\frac{s\cdot
\OPT}{s+1}\geq (s+1-1-\frac{s}{s+1})\cdot
\OPT=\frac{s^2+s-s}{s+1}\cdot \OPT=\frac{s^2\cdot \OPT}{s+1}$.
If $\mu < \frac{s}{s+1}$, we have $\OPT=\frac{\mu}{s}$. Thus
 $\sigma_1(i') \geq s\cdot \OPT
-\frac{s\cdot \OPT}{s+1}\geq \frac{s^2\cdot \OPT}{s+1}$.


This implies that $\min\{\frac{\sigma_1(i)}{s},\sigma_2(i)\}\geq
\min\{\frac{\sigma_1(i')}{s},\sigma_2(i')\} \geq
\frac{s\cdot\OPT}{s+1}$, where $i$ is the partition that SSNC
chooses for speed $s$.
If $\sigma_1(i)\geq \sigma_2(i)$, then the sets of jobs are not
resorted, and $M_1$ (resp.~$M_2$) receives a total of
$\sigma_1(i)$ (resp.~$\sigma_2(i)$), so we are done.
Otherwise, $M_1$ receives a load of
$\frac{\sigma_2(i)}{s}\geq \frac{\sigma_1(i)}{s}\geq \frac{s\cdot
\OPT}{s+1}$ and $M_2$ receives a load of ${\sigma_1(i)}\geq
\frac{\sigma_1(i)}{s}\geq \frac{s\cdot \OPT}{s+1}$.

\vspace{10pt}
\noindent
The case $s<\phi$ is deferred to the appendix.
For the case $s<\phi$, consider several cases. \Xomit{ Note that the
best possible job allocation (which might not be possible,
depending on the instance) is one where
$\sigma_1(i):\sigma_2(i)=s_1:s_2=s$. This means the largest job
set has size $s/(s+1)$ and the smallest $1/(s+1)$.}
In the sequel, if $s=1$, we consider an optimal assignment whose
work on $M_1$ is no smaller than its work on $M_2$. Note that
$M_1$ is always assigned $\max\{\sigma_1(i),\sigma_2(i)\}\geq
\frac 12$ by the algorithm. Since $\OPT\leq \frac{1}{s+1}$, an
optimal algorithm assigns at most $\frac{s}{s+1}$ to $M_1$ and we
get a ratio of $\frac{2s}{s+1}<1+\frac s{s+1}$. Thus $M_1$ gets
sufficient load. Let $i$ indicate the partition which is chosen by SSNC.

Suppose first that there exists a job of size at least $\frac
23$. Clearly, this is the first job and it belongs to the first set
found by SSNC, which has a larger size than the second set.
Also, for all other jobs $i\geq 2$ we have $p_i \leq \frac 13$.
Therefore $\sigma_1(i)\geq \frac 23$ and since $\OPT<1$, $M_1$
gets sufficient load.
If $i=1$, we are done since in the optimal assignment, the work
on $M_2$ is at most $\sigma_2(1)=1-p_1$.
Otherwise, $i\geq 2$. Using
(\ref{eq:tail}) we have
$\sigma_2(i)\geq(\sigma_1(i)-p_i)/s\geq(2/3)/s$ and thus
$\sigma_2(i)/\OPT\geq\frac 2{3s}/\frac 1{s+1}=\frac{2s+2}{3s}\geq\frac 2 3
\geq 1+\frac s{s+1}$.

Now suppose all jobs have size less than $2/3$.
If $p_{i}\leq1/3$ (and thus $p_{i+1}\leq \frac 13$ as well), we get from
(\ref{eq:head}) that
$\sigma_2(i)-p_{i+1}=1-\sigma_1(i)-p_{i+1}\leq \sigma_1(i)/s$,
which implies $\sigma_1(i)(s+1)\geq s(1-p_{i+1})\geq \frac{2s}{3}.$ Further, we get from
(\ref{eq:tail}) that
$(1-\sigma_1(i))s\geq \sigma_1(i)-p_i$, implying
$\sigma_1(i)\leq (s+p_i)/(s+1)$
and therefore
$\sigma_2(i)=1-\sigma_1(i)\geq(1-p_i)/(s+1)\geq2/(3s+3)$.
Thus $\min\{\sigma_1(i),\sigma_2(i)\} \geq\frac {2}{3(s+1)}\geq
\frac 2 3 \OPT\geq(1+\frac s{s+1})\OPT$.

If $p_{i}>1/3$, but $p_1 <\frac 23$, we have $i=1$ or $i=2$, since there are at most two
jobs larger than $\frac 13$.
If $i=1$, we have $\min\{\sigma_1(1),\sigma_2(1)\}=\min\{p_1,1-p_1\} >\frac 13 \geq\frac 2 3 \OPT\geq(1+\frac s{s+1})\OPT$.
If $i=2$, then $p_1>\frac 13$, and by (\ref{eq:tail}) we have
$\sigma_2(2)\geq \frac{\sigma_2(1)-p_2}{s}=\frac{p_1}{s}$.
We have $1=p_1+p_2+\sigma_2(2)\leq 2p_1+\sigma_2(2)\leq (2s+1)\sigma_2(2).$
Therefore $\OPT/\sigma_2(2)\leq \frac 1{s+1}/\frac1{2s+1}=
1+\frac s{s+1}$.
\end{proof}

It follows that on two machines, SSNC is better than SNC in general. However,
the following lemma shows that SNC is better than SSNC for $s\leq1+\sqrt2$.

\begin{lemma}
\label{lem:ssnc-lb}
The approximation ratio of SSNC is not better than
$\min\{1+\frac{s}{s+1},1+\frac 1 s\}$ on two machines.
\end{lemma}
\begin{proof}
Suppose $s\leq\phi$.
Consider the following input instance for some $\eps>0$:
jobs of size
$\frac s{2s+1},\frac s{2s+1}-\eps$, and many small jobs of total size
$1-\frac{2s}{2s+1}+\eps$. It is always possible to distribute these
jobs in a ratio of $s:1$, so the optimal cover is $1/(s+1)$.
For any $0<\eps<\frac s{2s+1}$, SSNC will combine the first two
jobs on the fast machine, and on the slow machine it will have a load of only
$1-\frac{2s}{2s+1}+\eps = \frac 1{2s+1}+\eps$.
Taking $\eps\to0$, this shows that for $s\leq\phi$, the approximation
ratio of SSNC is not better than $\frac1{s+1}/\frac1{2s+1}=\frac{2s+1}{s+1}$.

Now suppose $s>\phi$. In this case we use the jobs
$\frac{s^2}{(s+1)^2}-\eps, \frac1{s+1}+\eps$, and
$\frac{s}{(s+1)^2}$. These jobs are in order of decreasing size if $s>\phi$.
Again SSNC puts the first two jobs on the fast
machine, and has a cover of only $\frac{s}{(s+1)^2}$. The optimal assignment
is to combine the first and third jobs on the fast machine for a cover
of $\frac1{s+1}-\frac{\eps}{s}$.
\end{proof}


In the sequel, we show that SSNC or simple adaptations of it are
not monotone on more than two machines. In our examples we use a
small number of machines. The examples can be extended to a
larger number of machines by adding sufficiently many very large
jobs. We analyze an exponential version of SSNC that checks all
valid partitions of the sorted job list into $m$ consecutive sets.
Denote the sums of these sets by $X_1,\ldots,X_m$. Then SSNC
outputs the partition which maximizes $\min_{1\leq i\leq
m} \{ \frac {X_i}{ s_i}\}$.

Let $a>\sqrt{2}$. We use a job set which consists of five jobs of
sizes $a^3, a^3-1, a^2-1, a^2-1, 1$. There are three machines of
speeds $a^2,a,1$.

Running SSNC results in the sets $\{a^3\}, \{a^3-1\}, \{a^2-1,
a^2-1, 1\}$ for a cover of $a$. It is easy to see that changing
the first set into $\{a^3,a^3-1\}$ so that the load on the fastest
machine becomes strictly larger than $a$ results in a second set
$\{a^2-1,a^2-1\}$ and the third machine gets a load which is too
small.

Assume now the speed of fastest machine decreases from $a^2$ to
$a$. SSNC finds the sets  $\{a^3\}, \{a^3-1,a^2-1\}, \{a^2-1,1\}$
for a cover of $a^2$. So the size of the largest set can increase
(in this case, from $a^3$ to $a^3+a^2-2$) if the fastest machine
slows down.

This example shows that not only the above algorithm is not
monotone, but also a version of it which rounds machine speeds to
power of $a$. In previous work, machine speeds were rounded to
powers of relatively large numbers (e.g., $2.5$ in
\cite{AnAzSo05}). Thus it seems unlikely that  rounding machine
speeds to powers of some number smaller than $\sqrt{2}$ would
give a monotone algorithm.

Another option would be to round job sizes. We show that this
does not work either.

\begin{lemma}
The algorithm which rounds job sizes to powers of some value $b>\phi$
and then applies SSNC is not monotone for two machines.
\end{lemma}
\begin{proof}
Let $a$ be
a number such that $b <a <b+1 $. This is a constant used to define
machine speeds (the same example may be used to show that the
combination of rounding both machine speeds and job sizes is not
monotone either, since rounding speeds into powers of $a$ would
leave the speeds unchanged).
We consider the following problem instance with two machines and
five jobs. The speeds of both machines are $a$ initially, and the
job sizes are $(1+\eps)b,b,b,1$, where we take $\eps<1/b$.

Our algorithm sees the job sizes as $b^2,b,b,1$ and initially
places $b^2$ on machine 1 and the remaining jobs on machine 2.
Note that putting the first job of size $b$ also on machine 1
only gives a cover of $(b+1)/a$, whereas the first option gives
$b^2/a$ (and $b>\phi$). The algorithm then uses the actual job
sizes (which it needs to do in order to resort the job sets
accurately), and puts only the job of size $(1+\eps)b$ on the
second machine.

Now the speed of machine 2 decreases from $a$ to 1. The new job
sets are $\{b^2,b\}, \{b,1\}$, to get a (rounded) cover of
$(b^2+b)/a>b$. This hold since $(b^2+b)/a<b+1$. Keeping the old
sets would give only a cover of $b^2/a < b$. Taking the sets
$\{b^2,b,b\}$ and $\{1\}$ would give only a cover of $1$.
However, this means that the actual size of the first set is now
$(2+\eps)b$, whereas the size of the second set is $b+1$, which
is less. So the size of the smallest set is now $b+1$, which is
larger than before ($(1+\eps)b$), so the work on machine 2
increases although its speed decreased.
\end{proof}

Assume that we round job sizes to powers of $b>\phi$. Let $a$ be
a number such that $b <a <b+1 $. This is a constant used to define
machine speeds (the same example may be used to show that the
combination of rounding both machine speeds and job sizes is not
monotone either, since rounding speeds into powers of $a$ would
leave the speeds unchanged).
We consider the following problem instance with two machines and
five jobs. The speeds of both machines are $a$ initially, and the
job sizes are $(1+\eps)b,b,b,1$, where we take $\eps<1/b$.

Our algorithm sees the job sizes as $b^2,b,b,1$ and initially
places $b^2$ on machine 1 and the remaining jobs on machine 2.
Note that putting the first job of size $b$ also on machine 1
only gives a cover of $(b+1)/a$, whereas the first option gives
$b^2/a$ (and $b>\phi$). The algorithm then uses the actual job
sizes (which it needs to do in order to resort the job sets
accurately), and puts only the job of size $(1+\eps)b$ on the
second machine.

Now the speed of machine 2 decreases from $a$ to 1. The new job
sets are $\{b^2,b\}, \{b,1\}$, to get a (rounded) cover of
$(b^2+b)/a>b$. This hold since $(b^2+b)/a<b+1$. Keeping the old
sets would give only a cover of $b^2/a < b$. Taking the sets
$\{b^2,b,b\}$ and $\{1\}$ would give only a cover of $1$.
However, this means that the actual size of the first set is now
$(2+\eps)b$, whereas the size of the second set is $b+1$, which
is less. So the size of the smallest set is now $b+1$, which is
larger than before ($(1+\eps)b$), so the work on machine 2
increases although its speed decreased.

\section{Acknowledgment}

The authors would like to thank an anonymous referee who pointed out
an error in an earlier version of our approximation scheme in
Section \ref{sec:ptas}, 
another referee who helped improve the presentation,
and Motti Sorani for helpful discussions.


\end{document}